%% file: paper.tex
\documentclass[conference,10pt]{IEEEtran}

\usepackage{amsmath, mathrsfs}

\usepackage{stmaryrd}
\usepackage{relsize}

\usepackage{cite}
\usepackage{amsfonts}
\usepackage{amssymb}
\usepackage{color}
\usepackage{multirow}
\usepackage{graphicx}
\usepackage{relsize}
\usepackage{multirow}
\usepackage{multicol}
\usepackage{caption}
\usepackage{subcaption}
\usepackage[numbers,sort&compress]{natbib}
\usepackage{balance}
\usepackage{arydshln}
\allowdisplaybreaks

\usepackage{amsfonts}
\usepackage{amssymb}
\usepackage{graphicx}
\usepackage{color}
\usepackage{multirow}
\usepackage{graphicx}

\usepackage{caption}
\usepackage{subcaption}
\usepackage[numbers,sort&compress]{natbib}

\usepackage{algorithm}
\usepackage[noend]{algpseudocode}

\usepackage{lipsum}

\newcommand\blfootnote[1]{%
	\begingroup
	\renewcommand\thefootnote{}\footnote{#1}%
	\addtocounter{footnote}{-1}%
	\endgroup
}

\newcommand{\subparagraph}{}
\usepackage[compact]{titlesec}
\titlespacing*{\section}{15pt}{1\baselineskip}{0.9\baselineskip}

\allowdisplaybreaks
\newcommand{\myhash}{%
  {\settoheight{\dimen0}{C}\kern-.05em\, \resizebox{!}{\dimen0}{\raisebox{\depth}{\#}}}}

\usepackage{multicol}


\input{symbols.tex}

\newtheorem{assumption}[theorem]{{\bf Assumption}}

\usepackage{float}
\usepackage{afterpage}
\usepackage{balance}

\usepackage{geometry}
\newtheorem{observation}{\textbf{ Observation}}

\newtheorem{remark}{{\bf Remark}}
\def\herm{{\sfH}}

\def\cg{{\clC\clN}}

\input{macros.tex}

\graphicspath{ {./Figures/} }

\begin{document}
\title{WiFi-Based Channel Impulse Response Estimation and Localization via Multi-Band Splicing}
\author{\IEEEauthorblockN{Mahdi Barzegar Khalilsarai\IEEEauthorrefmark{1}, Benedikt Gross\IEEEauthorrefmark{2},
Stelios Stefanatos\IEEEauthorrefmark{3}, Gerhard Wunder\IEEEauthorrefmark{2}, and Giuseppe Caire\IEEEauthorrefmark{1}}
\vspace{-4mm}\\
$^{\ast}$Communications and Information Theory Group, Technische Universit\"{a}t Berlin,\\
%~\IEEEauthorrefmark{3}Qualcomm Technologies, Inc.\\
$^{\dagger}$Heisenberg Communications and Information Theory Group, Freie Universit\"{a}t Berlin,\\
\IEEEauthorrefmark{3}Qualcomm Technologies, Inc.\\
Emails: $\{$m.barzegarkhalilsarai, caire$\}$@tu-berlin.de, $\{$benedikt.gross,  g.wunder$\}$@fu-berlin.de,\\ sstefana@qti.qualcomm.com}
\maketitle
%%%%%%%%% Abstract%%%%%%%%%%%
%%%%%%%%%%%%%%%%%%%%%%%%%%%%
\begin{abstract}
Using commodity WiFi data for applications such as indoor localization, object identification and tracking and channel sounding has recently gained considerable attention. We study the problem of channel impulse response (CIR) estimation from commodity WiFi channel state information (CSI). The accuracy of a CIR estimation method in this setup is limited by both the available channel bandwidth as well as various CSI distortions induced by the underlying hardware. We propose a multi-band splicing method that increases channel bandwidth by combining CSI data across multiple frequency bands. In order to compensate for the CSI distortions, we develop a per-band processing algorithm that is able to estimate the distortion parameters and remove them to yield the ``clean" CSI. This algorithm incorporates the \textit{atomic norm denoising} sparse recovery method to exploit channel sparsity. Splicing clean CSI over $M$ frequency bands, we use \textit{orthogonal matching pursuit} (OMP) as an estimation method to recover the sparse CIR with high ($M$-fold) resolution. Unlike previous works in the literature, our method does not appeal to any limiting assumption on the CIR (other than the widely accepted sparsity assumption) or any ad hoc processing for distortion removal. We show, empirically, that the proposed method outperforms the state of the art in terms of localization accuracy.
\end{abstract}
\vspace{-1mm}
%%%%%%%%% Introduction%%%%%%%%%%%
%%%%%%%%%%%%%%%%%%%%%%%%%%%%
\section{Introduction}
\begin{figure}[t]
	\centering
	\includegraphics[ width=0.35\textwidth]{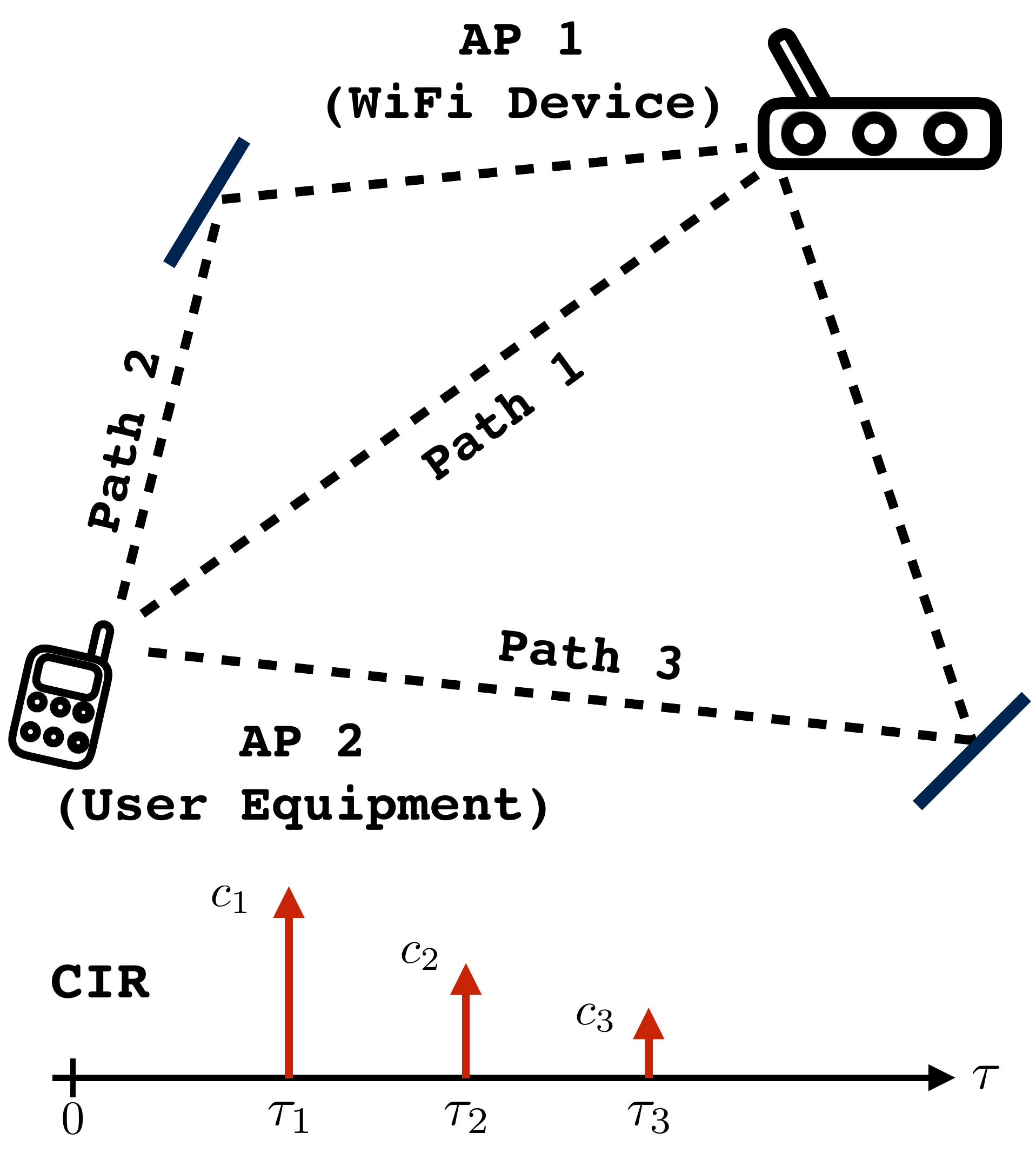}
	\caption{An example schematic of two communicating APs.}
	\label{fig:WiFi}
\end{figure}

\blfootnote{\IEEEauthorrefmark{3}The work of Stelios Stefanatos was performed while he was with the Freie Universit\"{a}t Berlin.}
Using commodity WiFi devices for purposes other than classical wireless communication has sparked considerable interest in the recent years, especially with the advent of the Internet of Things (IoT) technology \cite{li2011applications}. The reason is that WiFi devices are ubiquitous, operate at a low cost and their acquired data provide highly useful information about the environment. Prominent examples of applications that make use of commodity WiFi include indoor localization and ranging \cite{yang2015wifi,vasisht2016decimeter,khalilsarai2019wifi}, channel sounding \cite{maas2011channel}, object tracking \cite{qian2018widar2}, human activity recognition \cite{shen2019wirim}, etc. These applications rely on the channel state information (CSI) of a communication link between a pair of transmitter-receiver access points (APs). In general, the CSI refers to the received signal at the device in time, frequency and space or a combination thereof. As an example, in a single-antenna WiFi device, the CSI is equivalent to the \textit{channel frequency response} (CFR) across the used radio bands. The CFR reveals important information about the propagation environment. In particular, in a multi-path channel, the signal travels through distinct paths each with different attenuation, scattering, and diffraction properties, and with different ``travel times". These characteristics are formally represented by the \textit{channel impulse response} (CIR), which is a signal defined over the (propagation) delay domain (see Fig. \ref{fig:WiFi}) and is related to the CFR via the Fourier transform \cite{tse2005fundamentals}. The CIR can be directly used in any of the applications stated above. Its first delay component is equivalent to the channel Time-of-Flight (ToF), the time taken by the signal to travel from the transmitter to the receiver via a Line-of-Sight (LoS) path and can be used for localization \cite{vasisht2016decimeter}. The CIR also shows the power delay profile, which can be used for channel sounding purposes \cite{xie2018precise}. In addition, comparing the CIR at different times is applicable in recognizing human gesture and activity \cite{shen2019wirim}. 

The discussion above shows that meaningful application of WiFi CSI data hinges upon an accurate estimation of the CIR. A major challenge for achieving this goal is that, limited channel bandwidth (BW) results in a \textit{low-resolution} CIR estimate, due to the well-known time (delay)-frequency uncertainty principle. By a low resolution, we mean that the smaller the channel BW is, the harder it is to distinguish between two adjacent delay components in the CIR, even in the absence of noise. Conversely, by increasing the BW, we achieve higher resolutions and minute details of the CIR can be revealed. For example, a typical WiFi frequency band spans $20$ MHz of BW. This is roughly equivalent to a resolution of $\Delta \tau = 1/\text{BW}= 50$ ns in the delay domain, which means that two paths can be discriminated if the difference of their delays is larger than $50$ ns. Multiplied by the speed of light $\nu \approx 3\times 10^8$, this implies that the distance taken by the two paths must be larger than $\Delta d= \nu \Delta \tau=15$ m. In many practical scenarios (e.g. in small indoor environments), such condition is violated. A solution that has been recently studied is increasing the BW via \textit{multi-band splicing}, which is a technique that merges CFR samples across multiple bands to obtain high-resolution CIR estimates.

However, splicing CFR samples over multiple bands faces a major challenge due to hardware imperfections. The band carrier frequencies on the transmitter and receiver sides are not precisely the same and the discrepancy between the two varies from band to band \cite{chiueh2012baseband}. In addition, detecting a packet at the receiver introduces a delay that is added to the natural channel ToF \cite{vasisht2016decimeter}. Besides, frequency hopping across different bands causes a random phase offset in the CFR samples since the \textit{phase-locked loop} (PLL) in charge of generating the band carrier frequency starts from a random initial phase. The combined effect of these hardware imperfections results in a phase distortion in the CFR samples, which needs to be treated in implementing any reasonable multiband splicing method. As will be discussed in section \ref{sec:sys_set}, this phase distortion follows a piece-wise linear rule: it is linear with fixed unknown parameters per band, while the parameters vary from band to band. Several works in the literature have dealt with the problem of multi-band splicing. The Chronos system was proposed in \cite{vasisht2016decimeter}, which uses the squared CFR samples corresponding to the carrier frequency in each band to estimate the ToF via a compressed sensing sparse recovery method. This method is only designed to estimated the ToF and can not estimate the CIR. It also does not work in scenarios where only a few number of bands are used, since in this case the number of measurements is too small for any compressed sensing method to be useful. Spectrum splicing for ToF estimation was also studied in \cite{xie2018precise} and \cite{zhuo2017perceiving}, where heuristic methods were proposed with overall performances inferior to that of \cite{vasisht2016decimeter}. ToneTrack \cite{xiong2015tonetrack} is another indoor localization system which combines channel measurements across multiple bands to increase resolution. However, it is not designed for use with commodity WiFi data and therefore it is not clear how it can handle the existent signal phase distortions in WiFi CSI.
\subsection{Our Contribution}
Our proposed method consists of three steps, devised to address the aforementioned challenges: first, a per-band processing algorithm that estimates the linear phase distortion parameters and removes them from the CFR sample; a process that we shall call ``cleaning". This step involves atomic norm denoising \cite{bhaskar2013atomic}, which exploits channel sparsity to identify the phase distortion components over the continuous domain and without assuming them to be included in a discrete set of plausible values. Secondly, we use the clean CFR samples gathered from all bands to obtain a high-resolution estimate of the CIR. This step is implemented via the greedy orthogonal matching pursuit (OMP) sparse recovery method that exploits signal sparsity. The CIR obtained at this step is a ``relative" one, i.e. it contains ambiguities with respect to a shift in the delay domain as well as a global phase shift in the coefficients. As a final step we employ a handshaking protocol between the transmitter and the receiver to obtain extra information about the channel. This information is then used to form a likelihood function, which we maximize with respect to the delay and phase shift parameters and resolve the due ambiguities. 

\textbf{Notation:} Scalars are denoted by simple alphabet letters, while vectors and matrices are bold-faced. For an integer $M$, we define $[M]:= \{ 1,2,\ldots,M \}$. We also define the set of integers $\Nc$ as $\Nc = \{ -\tfrac{N-1}{2},\ldots,\tfrac{N-1}{2} \}$, where $N$ is an odd integer. The Fourier transform of a signal $h (\tau)$ at a point $f$ is denoted by $\Fc \{ h(\tau) \}|_{f}=\int h (\tau) e^{-j2\pi f \tau} d\tau$. For a matrix $\Xm$, $\Xm^\transp$ denotes transpose, $\Xm^\herm$ denotes Hermitian transpose, and $\Xm^\dagger = (\Xm^\herm \Xm )^{-1}\Xm^\herm$ denotes Moore-Penrose pseudo-inverse. 
%%%%%%%%% System Setup%%%%%%%%%%%
%%%%%%%%%%%%%%%%%%%%%%%%%%%%

\section{System Setup}\label{sec:sys_set}
We assume two \textit{access points} (APs) communicating with each other using OFDM signaling over $M$ frequency bands. Each band includes $N$ subcarriers indexed via the integer set $\Nc$, over which the pilot symbols are transmitted. The pilot signal at the receiver side can (ideally) be written as \cite{tse2005fundamentals}
\begin{equation}\label{eq:pilot_sig}
y [m,n] = H[m,n] S_{m,n} +z [m,n],~m\in [M],\, n\in \Nc,
\end{equation}
where $S_{m,n}=1$ is the symbol transmitted over subcarrier $f_{m,n}$ ($n$-th subcarrier of band $m$) and assumed to be equal to one for simplicity, $H[m,n] $ is the CFR sample at the same subcarrier and $z [m,n]\sim \cg (0,1/\SNR)$ denotes additive white Gaussian noise (AWGN). The CFR is related to the CIR via a Fourier transform. Suppose there exist $K$ scatterers in the propagation environment. Then the CIR is given as 
\begin{equation}\label{eq:CIR}
h(\tau) = \sum_{k=1}^K c_k \delta (\tau - \tau_k),
\end{equation}
where $\delta (\cdot)$ denotes Dirac's delta function, $\tau_k\in [0,1/f_s)$ and $c_k\in \bC$ are the delay and gain associated with path $k$, respectively, and where $f_s$ denotes subcarrier spacing. Note that the path delay and gain parameters are independent of the frequency band. The CFR samples are then given as
\begin{equation}\label{eq:CFR_CIR}
H[m,n] = \Fc \{ h(\tau) \}|_{f=f_{m,n}} = \sum_{k=1}^K c_k e^{-j2\pi f_{m,n}\tau_k},
\end{equation}
for $~m\in [M],\, n\in \Nc$. The pilot signal in a WiFi device is subject to several phase distortions, caused by hardware imperfections. These distortions result in the phase term
\begin{equation}\label{eq:affine_phase}
\begin{aligned}
\psi [m,n] &= -2\pi  (\delta_m nf_s + \phi_m),~m\in [M],\, n\in \Nc.
\end{aligned}
\end{equation} 
The parameter $\delta_m\in [0,1/f_s)$ represents receiver timing offset due to the packet detection delay (PDD) and receiver sampling frequency offset (SFO). Besides, the phase offset term $\phi_m\in [0,1)$ represents the effect of the random phase offset introduced by the PLL when switching channel bands, in addition to the accumulated phase offset across the slots due to the carrier frequency offset (CFO) between transmitter and receiver \cite{zhuo2017perceiving,vasisht2016decimeter,shen2019wirim}. 
%Besides, $\phi_m\in [0,1)$ represents a constant phase offset that stems from a discrepancy between transmitter and receiver carrier frequencies - also known as carrier frequency offset (CFO)- plus the random phase offset cause by the PLL \cite{zhuo2017perceiving,vasisht2016decimeter,shen2019wirim}.
 Notice that $\delta_m$ and $\phi_m$ differ from band to band, such that $\psi[n,m]$ is a piecewise linear function (on each band it is a linear function of the subcarrier index with different slope and constant terms). 
%As is explicitly denoted, these parameters generally differ from band to band which is why the phase distortion term is piece-wise linear. 

%Using \eqref{eq:pilot_sig}, the phase-distorted pilot signal can be written as
The actual received pilot signal including the unavoidable phase distortion is given by 
\begin{equation}\label{eq:pilot_sig_phase_dis}
\begin{aligned}
y [m,n] &=e^{j \psi [m,n]}\,  H[m,n]+z [m,n]\\
&  \hspace{-3mm}= e^{j \psi [m,n]}\sum_{k=1}^K c_k e^{-j2\pi f_{m,n}\tau_k}+z [m,n]\\
& \hspace{-3mm}= \sum_{k=1}^K c_k e^{-j2\pi (f_{m,0}\tau_k+\phi_m)} e^{-j2\pi nf_s(\delta_m+\tau_k)}+z [m,n],
\end{aligned}
\end{equation}
where we have used the fact that the subcarriers of a single band are equispaced with a spacing equal to $f_s$ and therefore we have $f_{m,n} = f_{m,0}+nf_s,~n\in \Nc$ with $f_{m,0}$ being the carrier frequency of band $m$. We denote the noise term as before by $z [m,n]$ since it is circularly symmetric and the multiplication of the phase-distortion term does not change its distribution.

The main problem can be posed as follows: given the noisy, phase-distorted pilot signals $y [m,n],~m\in [M],\,n\in \Nc$, estimate the CIR $h(\tau)$. Our proposed solution to this problem consists in the following steps:
\begin{enumerate}
	\item Estimate the distortion parameters $\{ \delta_m,\, \phi_m \}$ for all bands $m\in [M]$ and remove the phase distortion from the pilot signals.
	\item Splice clean pilot data over all bands to obtain a high-resolution CIR estimate.
	\item Use a \textit{hand-shaking} procedure between the APs to resolve ambiguities and estimate the ToF. 
\end{enumerate}
In the following we explain each of these steps in details. However, first we need to clarify what we mean by ``ambiguities" in step (3). 
\begin{observation}[Inherent Ambiguity]\label{obs:inherent_ambiguity}
	Suppose that a specific CIR $h^{(1)}(\tau) = \sum_{k=1}^K c_k^{(1)} \delta (\tau - \tau_k^{(1)})$ in combination with a specific set of phase distortion parameters $\delta_m^{(1)},\phi_m^{(1)},~m\in [M]$ and AWGN samples $z^{(1)}[m,n]$ generates pilot signals $y^{(1)}[m,n]$ according to \eqref{eq:pilot_sig_phase_dis}. Now, consider a second CIR $h^{(2)}(\tau) = \sum_{k=1}^K c_k^{(2)} \delta (\tau - \tau_k^{(2)}) $ and associated phase distortion parameters $\delta_m^{(2)},\phi_m^{(2)},~m\in [M]$ and the same AWGN samples $z^{(1)}[m,n]$. Let the parameters in the latter case be such that:
	\begin{equation}\label{eq:params}
	\begin{aligned}
	\tau_{k}^{(2)} & = 	\tau_{k}^{(1)}  - \bar{\delta},~~
	&&c_k^{(2)} = c_k^{(1)}e^{j2\pi \bar{\phi}},~\text{for all}~k,\\
	\delta_m^{(2)} & = \delta_m^{(1)}+\bar{\delta},~~
	&&\phi_m^{(2)} = \phi_m^{(1)}+\bar{\phi},~~\text{for all}~m,\\
	\end{aligned}
	\end{equation}
	where $\bar{\delta}$ and $\bar{\phi}$ are arbitrary values. Then it is easy to show that the generated pilot signals in the two cases coincide, i.e. $y^{(2)} [m,n] =y^{(1)} [m,n]$ for all $m,n$ for all $m,n$. Therefore, given the pilot signals, both sets of parameters in cases (1) and (2) are equally plausible. This shows that there exist two inherent ambiguities in estimating the CIR parameters: (a) an ambiguity in terms of a (circular) shift of the CIR over the delay domain and (b) an ambiguity in terms of a global phase shift of the CIR coefficients. Therefore, given only the phase-distorted pilot signals as in \eqref{eq:pilot_sig_phase_dis}, we are undecided about two scalar parameters, namely a delay shift and a phase shift. The delay shift is especially important here, as it effects the estimation of the ToF. In step (3) of our proposed method, we introduce a hand-shaking procedure to acquire some additional information about the CIR that helps resolve these ambiguities. 
\end{observation}

\begin{assumption}[Relative CIR]\label{assumption1}
	The observation above conveys an algorithmic implication. Without loss of generality (w.l.o.g), we can assume the CIR estimated from pilot measurements to always begin at $\tau=0$ and its first coefficient to have a zero phase. This is equivalent to shifting all CIR delay parameters by $\tau_1$ and subtracting the phase of the first coefficient $c_1$ from the phases of all coefficients. Mathematically, via this assumption we try to recover a ``relative" CIR of the form
	\begin{equation}\label{eq:h_0}
h_0 (\tau) = \sum_{k=1}^{K} \widetilde{c}_k \delta (\tau - \widetilde{\tau}_k),
\end{equation}
	where
%	\begin{equation}\label{eq:coeffs_delays_h0}
%	\begin{aligned}
$\widetilde{c}_k = c_k e^{-j\angle c_1},~
\widetilde{\tau}_k = \tau_k - \tau_1,$
%	\end{aligned}
%	\end{equation}
for $k=1,\ldots,K$ with $\angle c_1$ being the phase of the first CIR coefficient. Note that $\widetilde{\tau}_1=0$ and $\widetilde{c}_1=|c_1|$. Once $h_0 (\tau)$ is estimated, we can recover $\tau_1$ and $\angle c_1$ via the handshaking protocol.
\end{assumption}

\section{Removing Phase Distortion from Pilot Signals}
We propose a method to clean the pilot measurements by adopting the relative CIR model in \eqref{eq:h_0} and removing the phase distortion terms for each band separately. First, note that for each band $m$, the pilot signals $y[m,n],\, n\in \Nc$ represent the samples of a noisy mixture of $K$ complex sinusoids (see \eqref{eq:pilot_sig_phase_dis}). Using the relation $f_{m,n} = f_{m,0}+nf_s,~n\in \Nc$, the vector of pilot signals in band $m$ can be written as
\begin{equation}\label{eq:y_m}
\begin{aligned}
y [m,n] &= \sum_{k=1}^K c_k^{(m)} e^{-j2\pi n f_s \tau_{k}^{(m)}} +z[m,n] \\
& =: x [m,n]+z[m,n],
\end{aligned}
\end{equation}
where 
\begin{equation}\label{eq:coeffs_delays_m}
\begin{aligned}
	c_k^{(m)} &=c_k e^{-j2\pi (f_{m,0}\tau_k+\phi_m)},\\
	\tau_{k}^{(m)} &= \tau_k + \delta_m,
\end{aligned}
\end{equation}
 and we have denoted the signal part of $y[m,n]$ by $x[m,n]$. Our goal consists in estimating, for each band $m$, the set of parameters $\{ \tau_k^{(m)} \}_{k=1}^K$ and $\{ c_k^{(m)} \}_{k=1}^K$. Since the CIR is sparse ($K\ll N$), sparse recovery via compressed sensing is a natural choice for an estimation method. A powerful tool in this regard is the atomic norm denoising superresolution technique. Define the atoms $\av (\tau,\theta) \in \bC^{N}, ~ \tau \in [0,1/f_s],~ \theta\in [0,2\pi)$ with elements
%\begin{equation}\label{eq:a_atoms}
$[\av (\tau,\theta)]_n = e^{-j(2\pi n f_s \tau - \theta )},~n\in \Nc.$
%\end{equation}
The set of atoms above constructs a continuous dictionary $\Ac=\{ \av (\tau,\theta): \tau \in [0,1/f_s],\, \theta \in [0.2\pi) \}$. Now, let $\yv (m)=\left[  y[m,-\tfrac{N-1}{2}],\ldots,y[m,\tfrac{N-1}{2}] \right]^\transp \in \bC^N$ denote the vector of pilot samples in band $m$, $\xv (m)=\left[  x[m,-\tfrac{N-1}{2}],\ldots,x[m,\tfrac{N-1}{2}] \right]^\transp$ its signal component and $\zv (m)$ the associated AWGN vector. It follows that
\begin{equation}\label{eq:sig_part}
\xv (m) = \sum_{k=1}^K |c_k| \av (\tau_k^{(m)},\angle c_k^{(m)}),
\end{equation}
that is, $\xv (m)$ is a linear combination of a few elements of $\Ac$. We can define the atomic norm of a generic vector $\rv \in \bC^{N}$ over $\Ac$ is defined as \cite{tang2013compressed}
\begin{equation}\label{eq:atomic_norm}
\Vert \rv \Vert_{\Ac} = \underset{\underset{\theta_k\in [0,2\pi)}{w_k\ge 0, \, \tau_k\in [0,1/f_s] }}{\inf}\scalebox{1.3}{$\{$} \sum_{k}w_k \, : \, \rv = \sum_k w_k \av (\tau_k,\theta_k) \scalebox{1.3}{$\}$}.
\end{equation}
It is well known that the atomic norm promotes a sparse representation of $\rv$ in $\Ac$ and therefore is a suitable choice for estimating the sparse vector $\xv (m)$. If there were no noise, we would directly observe $\yv (m)=\xv (m)$ and the sparse recovery problem could be posed as \cite{tang2013compressed}:
\begin{equation}\label{eq:opt_1}
\underset{\widetilde{\xv}\in \bC^N}{\text{minimize}}\, \Vert \widetilde{\xv}\Vert_\Ac~~\text{subject to}~~ [\widetilde{\xv}]_n=[\yv (m)]_n,~n\in \Nc,
\end{equation}
which simply finds the vector that has the sparsest representation in $\Ac$ and whose elements are equal to the observed CFR samples of band $m$. In the problem in hand, however, we have access only to the noisy pilot observations $\yv (m)=\xv (m)+ \zv (m)$. In this case, the \textit{atomic norm denoising} method was proposed in \cite{bhaskar2013atomic} which solves the following regularized convex problem:
\begin{equation}\label{eq:opt_2}
\underset{\widetilde{\xv}\in \bC^N}{\text{minimize}}\, \frac{1}{2} \Vert \widetilde{\xv}-\yv (m) \Vert^2 +\lambda_m \Vert  \widetilde{\xv}\Vert_{\Ac},
\end{equation}
where $\lambda_m>0$ is a suitable regularization scalar. Fortunately, the atomic norm denoising optimization problem has an equivalent SDP form as \cite{bhaskar2013atomic}
\begin{equation}\label{eq:opt_SDP}
\begin{aligned}
&\underset{t,\uv,\widetilde{\xv}}{\text{minimize}}&&\frac{1}{2} \Vert \widetilde{\xv}-\yv (m) \Vert^2 + \frac{\lambda_m}{2}(t+[\uv]_1)\\
&\text{subject to} &&\begin{bmatrix}
T(\uv)&\widetilde{\xv}\\
\widetilde{\xv}^\herm & t
\end{bmatrix}\succeq \mathbf{0},
\end{aligned}
\end{equation}
where $T(\uv)$ is the Toeplitz Hermitian matrix with $\uv$ as its first column. The solution of \eqref{eq:opt_SDP} for $\widehat{\xv}$ is a vector $\widehat{\xv} (m) $ that has the following representation 
\[ \widehat{\xv} (m) = \sum_{k} |\widehat{c}_k^{(m)}| \av (\widehat{\tau}_k^{(m)},\angle \widehat{c}_k^{(m)}) = \sum_{k}\widehat{c}_k^{(m)} \av (\widehat{\tau}_k^{(m)},0).  \]
We are interested in the parameters $ \widehat{\tau}_k^{(m)},\, \widehat{\theta}_k^{(m)}$, as they contain the phase distortion data. The harmonic parameters $\widehat{\tau}_k^{(m)}$ can be obtained by solving the dual problem of \eqref{eq:opt_SDP} and finding those points in which the dual polynomial has a maximum modulus (details are omitted due to a lack of space; see \cite{bhaskar2013atomic}). Denote the estimated support of $\widehat{\xv}(m)$ over the delay domain by $\widehat{\tau}_k^{(m)},\, k=1,\ldots,\widehat{K}$. The coefficients $\{\widehat{c}_k^{(m)}\}_{k=1}^{\widehat{K}}$ are computed as $\widehat{\cv}^{(m)} =  [\widehat{c}_1^{(m)},\ldots,\widehat{c}_{\widehat{K}}^{(m)}]^\transp = \Am^{(m)\, \dagger} \yv (m)$, where $\Am^{(m)} := \left[ \av (\widehat{\tau}_1^{(m)},0),\ldots,\av (\widehat{\tau}_{\widehat{K}}^{(m)},0) \right]$.
\begin{remark}
	The optimal value for the regularization parameter $\lambda_m$ depends on the dual norm of the noise term, and for the AWGN model is derived as \cite{bhaskar2013atomic}
	\begin{equation}\label{eq:lambda_val}
	\lambda_m =\frac{1+1/\log N}{\sqrt{\SNR}}\sqrt{N\log N+N \log \left(4\pi \log  N\right)}.
	\end{equation}
Hereafter, we use this value for the regularization parameter for solving \eqref{eq:opt_2}.
\end{remark}

\subsection{Removing phase distortions}
Recall that via Assumption \ref{assumption1}, we postulate (w.l.o.g) the CIR to be equal to the relative CIR $h_0 (\tau)$ in \eqref{eq:h_0} with parameters $\widetilde{c}_k,\, \widetilde{\tau}_k,\, k=1,\ldots,K$ where $\angle \widetilde{c}_1 = 0$ and $\widetilde{\tau}=0$.  Assuming this and since $\widehat{\xv}(m),\,m\in [M]$ are estimated from the phase-distorted measurements, from \eqref{eq:coeffs_delays_m} we have that $\widehat{\tau}^{(m)}_1$ is an estimate of $\delta_m$ and $\angle \widehat{c}_1^{(m)}$ is an estimate of $-2\pi \phi_m$. In other words, the estimates of the distortion parameters for band $m$ are given as:
\begin{equation}\label{eq:est_phase_dist}
\begin{aligned}
 \widehat{\delta}_m &= {\sf est}(\delta_m) && = \widehat{\tau}^{(m)}_1, \\
\widehat{\phi}_m &= {\sf est}(\phi_m) && = -\frac{\angle \widehat{c}_1^{(m)}}{2\pi}.
\end{aligned}
\end{equation}
Therefore, we can remove the phase distortions by calculating the clean pilot CFR samples as 
\begin{equation}\label{eq:cleaned_y_samples}
\widetilde{y}[m,n] = e^{j 2\pi  (\widehat{\delta}_m nf_s + \widehat{\phi}_m)} y [m,n].
\end{equation}

\section{Multi-Band Splicing and CIR Estimation}
The idea of multi-band splicing is to combine the clean pilot measurements over all $M$ bands to estimate the CIR $h(\tau)$. The benefit of splicing several bands is that it increases the achievable resolution in estimating $h(\tau)$ by increasing the measurement bandwidth. While with a single band we have a resolution of $(\Delta \tau)_1 =1/Nf_s$ over the delay domain, using $M$ bands we can potentially achieve a resolution of $(\Delta \tau)_M= 1/MNf_s$ which is $M$-fold smaller (finer). 
\subsection{Estimating the Relative CIR $h_0 (\tau)$}
For band $m$, define the vector of band subcarriers as $\fv (m)= [f_{m,-\tfrac{N-1}{2}},\ldots,f_{m,\tfrac{N-1}{2}}]^\transp$ and the vector containing all subcarriers as $\fv = [\fv (1)^\transp,\ldots,\fv(M)^\transp]^\transp \in \bR^{MN}$. Similarly, we can define the vector of clean pilot measurements corresponding to the subcarriers as follows. Let $\widetilde{\yv}(m) = \left[\widetilde{y}[m,-\tfrac{N-1}{2}],\ldots, \widetilde{y}[m,\tfrac{N-1}{2}]\right]^\transp$ denote the clean measurements vector of band $m$ and $\widetilde{\yv} = [\widetilde{\yv}(1)^\transp,\ldots, \widetilde{\yv}(M)^\transp]^\transp\in \bC^{MN}$ as the multi-band spliced vector containing all CFR measurements. We can write the elements of $\tilde{\yv}$ as 
\begin{equation}\label{eq:y_tilde_h_0}
[\widetilde{\yv}]_i = \Fc \{ h_0 (\tau) \}|_{[\fv]_i } + [\widetilde{\zv}]_i,~i=1,\ldots,MN
\end{equation}
where $[\tilde{\zv}]_i$ represents the AWGN plus the error produced by the phase-distortion removal procedure (i.e. the error in estimating the parameters $\delta_m,\,\phi_m,~m\in [M]$). Since $h_0 (\tau)$ is sparse, a standard approach to estimating it from the frequency samples $\tilde{\yv}$ is using a compressed sensing method. Define a uniform grid of size $G$ over the delay domain as $\Gc = \{ 0,\frac{1}{G},\ldots,\frac{G-1}{G} \}/f_s$. Let $\Dm = [\dv (0),\ldots,\dv (G-1)] \in \bC^{MN\times G}$ be an overcomplete dictionary ($G\gg MN$) associated with $\Gc$, where each column $\dv (i) $ is defined as
\begin{equation}\label{eq:atom_def}
\scalebox{0.98}{$\dv (i) = \frac{1}{\sqrt{MN}}[e^{-j2\pi [\fv]_1 (\frac{i}{G})/f_s},\ldots,e^{-j2\pi [\fv]_{MN} (\frac{i}{G})/f_s}]^\transp \in \bC^{MN}$}
\end{equation}
for $i=0,1,\ldots,G-1$. Now, if the grid $\Gc$ is dense enough, we can approximate the vector form of \eqref{eq:y_tilde_h_0} as
\begin{equation}\label{eq:y_tilde_approx}
\widetilde{\yv} \approx \Dm \hv_0 +\widetilde{\zv},
\end{equation}
where $\hv_0 \in \bC^G$ is a discrete approximation for $h_0 (\tau)$. In practice, setting $G$ to a value of $G=2MN$ or $G=3MN$ yields a sufficiently dense grid. 

Given $\widetilde{\yv}$, we estimate the $\hv_0$ using the well-known \textit{orthogonal matching pursuit} (OMP) sparse recovery method \cite{cai2011orthogonal}. Using the model \eqref{eq:y_tilde_approx}, OMP can be seen as a greedy iterative algorithm that selects, at each iteration, a column of the dictionary $\Dm$ that has the highest correlation with the current residual and repeats this until a convergence condition is met. Then, the non-zero coefficients associated with each selected column to approximate the measurements vector $\widetilde{\yv}$ are computed by solving a simple least-squares problem. We refer the reader to \cite{cai2011orthogonal} for details. As for the stopping condition, we assume that the sparsity order of $\hv_0$ is given and halt the algorithm once the number of selected dictionary columns is equal to the sparsity order.

\begin{remark}
One may wonder why we do not use atomic norm denoising once more in this setting. The reason is twofold: first, the atomic norm has an SDP representation when we have uniformly obtained samples of the complex exponential mixture. The frequency samples limited to a single band are in fact taken over uniformly spaced subcarriers $f_{m,0}+n\, f_s$ for $n=-\frac{N-1}{2},\ldots,\frac{N-1}{2}$ so that the SDP form is guaranteed. In contrast, in general the elements of the multi-band subcarrier vector $\fv$ do not necessarily lie on a uniform grid. In this case a tractable formulation of the atomic norm is unknown. Secondly, even if the subcarriers lie on a uniform grid and the SDP form exists, solving the corresponding SDP is not desirable due to the high computational complexity of SDPs in large dimensions. Therefore, using a grid-based compressed sensing method such as OMP is more reasonable.   
\end{remark}

\subsection{Handshaking, Resolving Ambiguities and ToF Estimation}
By Observation \ref{obs:inherent_ambiguity} we know that with the given phase-distorted pilot signals, there always exists an ambiguity with respect to a circular shift of the CIR over the delay domain and a global phase shift in the coefficients. From this observation we concluded that, w.l.o.g, we can assume the relative CIR $h_0 (\tau)$ that has its first delay component at $\tau=0$ with a zero-phased coefficient (see \eqref{eq:h_0}). The relation between the relative and the true CIRs is as follows:
\begin{equation}\label{eq:h_0_h_true}
 h(\tau) = e^{j\angle c_1} h_0 (\tau -\tau_1).
\end{equation}
Therefore it remains to estimate $\tau_1$ and $c_1$ using extra information about the channel. This extra information is obtained via a handshaking procedure that was suggested in \cite{vasisht2016decimeter} and is explained in the following.

From the affine phase distortion model developed before, we know that the zero subcarrier in each band (band carrier frequency) is not polluted by the PDD and SFO phase errors, but only by the constant phase error term $\phi_m$ induced by the CFO and PLL phase offsets. This constant error term has the same absolute value but different signs on the transmitter-receiver APs (see \cite{vasisht2016decimeter} Eqs. (11) and (12)). Therefore, for each band $m$, we can write the pilot CSI at the two APs and on the center carrier frequency as
\begin{equation}
\begin{aligned}
y_{tx}[m,0] &=e^{j \phi_m}\,  H[m,0]+z_{tx} [m,0]\\
y_{rx}[m,0] &=e^{-j \phi_m}\,  H[m,0]+z_{rx} [m,0],
\end{aligned}
\end{equation}
where $z_{tx}[m,0] $ and $z_{rx}[m,0]$ denote noise terms at the transmitter and receiver sides, respectively. Notice the difference in the sign of the phase distortion terms on both sides. Also note that the CFR $H[m,0]$ is the same on both ends due to channel reciprocity \cite{tse2005fundamentals}. During a frequency-hopping procedure the transmitter sends packets to the receiver through which the receiver obtains the zero carrier measurements $y_{tx}[m,0],\, m=1,\ldots,M$. Therefore, at the receiver side we have both $y_{rx}[m,0]$ as well as $y_{tx}[m,0]$ for all $m$. Multiplying the two values for each $m$ we get
\begin{equation}\label{eq:measurement_multiplication}
y_{rx}[m,0]\, y_{tx}[m,0] = H[m,0]^2 + z'[m,0],~m\in [M],
\end{equation}
where $z'[m,0]:= e^{j \phi_m}\,  H[m,0] z_{rx}[m,0] + e^{-j \phi_m}\,  H[m,0]z_{tx}[m,0]+ z_{rx}[m,0] z_{tx}[m,0] $ denotes the signal and noise cross-terms, which especially in high-SNR regimes can be safely assumed to be of small value. Equation \eqref{eq:measurement_multiplication} gives the noisy squared CFR samples on the zero carrier per band. On the other hand, using the discrete estimate of the relative CIR $\widehat{\hv}_0$ obtained in the previous step and the relation \eqref{eq:h_0_h_true}, we can estimate $H [m,0]$ as a function of the unknown delay shift $\bar{\tau}$ and phase shift $\bar{\theta}$ parameters as 
\begin{equation}\label{eq:H_m_0_est}
\widehat{H}_{(\bar{\tau},\bar{\theta})}[m,0] = e^{j\bar{\theta}} e^{-j2\pi f_{m,0}\bar{\tau}} \sum_{i=1}^G\,  [\widehat{\hv}_0]_i\, e^{-j2\pi f_{m,0}(\frac{i-1}{G})/f_s}
\end{equation} 
for $m\in [M]$. Note that if the estimate of the relative CIR were exact, then we would have $\widehat{H}_{(\tau_1,\angle c_1)}[m,0] = H [m,0],\, m\in [M]$, i.e. the estimate coincides with the true CFR for parameters $\bar{\tau}=\tau_1$ and $\bar{\theta} = \angle c_1$. This observation implies that a natural way to estimate $\tau_1$ and $\angle c_1$ is to compare $\widehat{H}_{(\tau_1,\angle c_1)}[m,0]^2$ with $y_{rx}[m,0]\, y_{tx}[m,0] = H[m,0]^2 + z'[m,0]$ from \eqref{eq:measurement_multiplication} and minimizing their difference simultaneously for all $m\in [M]$. Formally, this can be cast as the following problem:
 
 \begin{equation}\label{eq:LS_problem}
\scalebox{0.91}{$\underset{(\bar{\tau},\bar{\theta})\in \Xim}{\text{minimize}}~ C(\bar{\tau},\bar{\theta})= \sum_{m=1}^M \left|  y_{rx}[m,0]\, y_{tx}[m,0] - \widehat{H}_{(\bar{\tau},\bar{\theta})}[m,0]^2 \right|^2.$}
 \end{equation}
 The variable domain $\Xim $ can be chosen in various ways. We have empirically observed that choosing a discrete set as
 \[ \Xim = \{ (\tau_i,\theta_i)\in [0,1/f_s)\times [0,2\pi),\, i=1,\ldots,I \} \]
 is sufficient. Then minimizing the cost $C(\bar{\tau},\bar{\theta})$ is a simple and fast minimization over a discrete set. Defining the optimal parameters as 
 \begin{equation}\label{eq:argmin}
(\tau^\star,\theta^\star) = \underset{(\bar{\tau},\bar{\theta})\in \Xim}{\arg \min}~ C(\bar{\tau},\bar{\theta}),
 \end{equation} 
the estimate of the true CIR is given by
\begin{equation}\label{eq:final_est}
\widehat{h}(\tau) = e^{j\theta^\star}\sum_{i=1}^G \, [\widehat{\hv}_0]_i\, \delta (\tau - \frac{i-1}{G}-\tau^\star).
\end{equation}
 In addition, the ToF is estimated as
 \begin{equation}\label{eq:ToF_est}
 \widehat{\text{ToF}} = \widehat{\tau}_1= \tau^\star.
 \end{equation}

\section{Simulation Results}
In this section, we compare our method to Chronos \cite{vasisht2016decimeter} in terms of localization accuracy. To this end, let us first define the average ranging estimation error as
\begin{equation}\label{eq:e_d}
e_d =|\tau_1 - \widehat{\tau}_1| \nu,
\end{equation}
where $\nu \approx 3\times 10^8$ is the speed of light. To generate random CIRs, we consider channels with $K=3$ delay taps. The maximum path distance is set to $d_{\max}=100~m$, which means that the longest path takes a total distance of $d_{\max}$ to reach the receiver, which is a reasonable choice for an indoor environment. This is equivalent to setting the maximum delay spread equal to $\tau_{\max}=d_{\max}/\nu \approx 333~ns$. The delay components are chosen uniformly at random in the interval $[0,\tau_{\max}]$. The path gains $\{c_k \}_{k=1}^K$ are generated as complex Gaussian random variables, with decreasing variance as $\sigma_k^2=4^{-k},\, k=1,\ldots,K$. This is a natural choice, since typically the path with a larger delay is attenuated more than a path with a smaller delay. We consider $M=16$ frequency bands each with $N=65$ subcarriers. With a subcarrier spacing of $f_s = 312.5$ kHz, each band occupies $(N-1)f_s=20$ MHz of bandwidth. As a proof of concept, we have considered half of the bands (8 bands) in the $(2,2.19)$ GHz range and the other half in the $(5,5.19)$ GHz range. Note that the $5$ GHz range is already in use by the IEEE 802.11a WiFi standard \cite{IEEEdoc1999}. For each band $m$, the phase distortion parameters are randomly generated as follows: we assume the PDD and SFO to cause a delay distortion of up to $960$ ns, so that $\delta_m$ is chosen uniformly at random (and independently across bands) in the range $[0,960]$ ns. Also the PLL and CFO induced constant phase distortion term $\phi_m$ is chosen uniformly at random (and independently across bands) in $[0,2\pi)$. We assume the signal-to-noise-ratio (SNR) to be equal to $\SNR = 20$ dBs. 

The Chronos method uses only the $M$ handshaking carrier frequency CFR samples in \eqref{eq:measurement_multiplication}, which are noisy versions of $\{ H[m,0]^2 \}$. Then, using the CIR sparsity assumption, it applies the \textit{basis pursuit denoising} (BPDN) algorithm to these measurements. BPDN can be formulated as
\[ \underset{\xv \in \bC^{G'}}{\text{minimize}} ~\Vert \xv \Vert_1~\text{subject to}~ \Vert \qv - \Fm \xv  \Vert \le \epsilon, \]
where $\qv = \left[y_{rx}[1,0]\, y_{tx}[1,0],\ldots,y_{rx}[M,0]\, y_{tx}[M,0]  \right]^\transp $, $\Fm \in \bC^{M\times G'}$ is the over-sampled Fourier matrix where $[\Fm]_{m,i} = \frac{1}{\sqrt{M}}e^{-j2\pi f_{m,0}\frac{i-1}{G'}/f_s},~m\in M,\, i\in [G']$ with $G'$ is typically chosen as a multiple of $M$. Also $\epsilon$ is an estimate on the $\ell_2$ norm of the noise plus cross-terms vector $\zv'= \left[ z'[1,0],\ldots,z'[M,0] \right]^\transp$ ($z'[m,0]$ is defined in \eqref{eq:measurement_multiplication}). Although this can not be done in practice, to be concrete, we set $\epsilon$ to be exactly equal to $\Vert \zv' \Vert$ in our simulations. The result is an sparse estimate of the CIR convoluted with itself $h(\tau)\ast h(\tau)$ (since $H[m,0]^2 = \Fc \{ h(\tau)\ast h(\tau) \}|_{f=f_{m,0}}$)and therefore, its first significant component is expected to be located at $2\tau_1$. Dividing this location by $2$ gives an estimate of $\tau_1$, i.e. the ToF. We do not further explain this method due to a lack of space and refer the reader to \cite{vasisht2016decimeter} for details.
\begin{figure}[t]
	\centering
	\includegraphics[ width=0.47\textwidth]{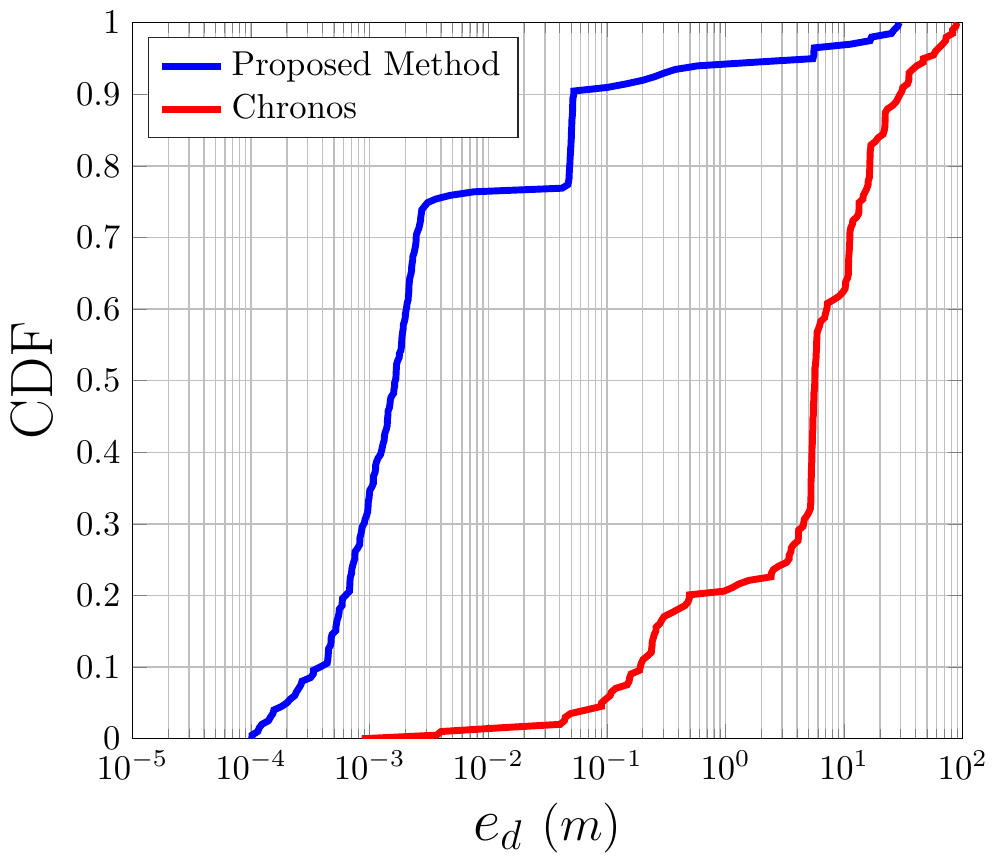}
	\caption{CDF of the ranging error for Chronos and our proposed method. Here we have $M=16$, $N=65$ and $\SNR =20$ dBs. }
	\label{fig:CDF}
\end{figure} 
Fig. \ref{fig:CDF} illustrates the ranging error CDF for our proposed method as well as for Chronos. The CDF is obtained after 200 Monte-Carlo simulations. As we can see, our proposed method achieves a ranging error that is 2 to 3 orders of magnitude smaller than Chronos: in 90 \% of the instances our method achieves a ranging error of $\approx 5$ cm or less, whereas the ranging error for Chronos is $\approx 34$ m or less. This figure also shows that in about 5 \% of the instances the ranging error of our method is 1 meters or larger. The reason for such outliers is that sometimes the phase distortion parameters are estimated poorly, either due to noise or two delay components being very close such that the atomic-norm denoising step fails to distinguish between them. In such cases, compensating for phase distortions induces an error in the estimation of the relative CIR and eventually the handshaking step fails to correctly identify the ToF. However, such errors happen rarely and do not effect the average performance of our method.   
\section{Conclusion}
We proposed a method for CIR estimation and indoor localization using commodity WiFi CSI data. This method performs multi-band CFR splicing and exploits channel sparsity to achieve high resolution CIR estimates. In order to compensate for inherent hardware-induce distortion in the CSI. we proposed a per-band processing based on atomic norm denoising which estimates the distortion parameters and remove them from the CSI samples. Via empirical simulations we showed that our method outperforms the state-of-the-art in terms of ranging error. 

\balance
{\small
	\bibliographystyle{IEEEtran}
	\bibliography{references}
}

\end{document}

%% file: symbols.tex
\usepackage{mathbbol}

\def\mindex#1{\index{#1}}

\def\sq{\hbox{\rlap{$\sqcap$}$\sqcup$}}
\def\qed{\ifmmode\sq\else{\unskip\nobreak\hfil
\penalty50\hskip1em\null\nobreak\hfil\sq
\parfillskip=0pt\finalhyphendemerits=0\endgraf}\fi\medskip}

\long\def\defbox#1{\framebox[.9\hsize][c]{\parbox{.85\hsize}{%
\parindent=0pt
\baselineskip=12pt plus .1pt      % STYLE
\parskip=6pt plus 1.5pt minus 1pt % CHANGES
 #1}}}

\long\def\beginbox#1\endbox{\subsection*{}%
\hbox{\hspace{.05\hsize}\defbox{\medskip#1\bigskip}}%
\subsection*{}}

\def\endbox{}

\newsavebox{\junk}
\savebox{\junk}[1.6mm]{\hbox{$|\!|\!|$}}

\def\bC{{\mathbb C}}

\def\bR{{\mathbb R}}

\def\sfH{{\sf H}}

\def\bfmath#1{{\mathchoice{\mbox{\boldmath$#1$}}%
{\mbox{\boldmath$#1$}}%
{\mbox{\boldmath$\scriptstyle#1$}}%
{\mbox{\boldmath$\scriptscriptstyle#1$}}}}

\def\bfmY{\bfmath{Y}}

\def\bfmhhaY{\bfmath{\hhaY}} %\widehat{\widehat{Y}}}}
\def\bfmhhaY{\hbox to 0pt{$\widehat{\bfmY}$\hss}\widehat{\phantom{\raise 1.25pt\hbox{$\bfmY$}}}}

\def\til={{\widetilde =}}

\def\clC{{\cal C}}

\def\clN{{\cal N}}

 \def\FRAC#1#2#3{\genfrac{}{}{}{#1}{#2}{#3}}

\def\ddtp{{\mathchoice{\FRAC{1}{d^{\hbox to 2pt{\rm\tiny +\hss}}}{dt}}%
{\FRAC{1}{d^{\hbox to 2pt{\rm\tiny +\hss}}}{dt}}%
{\FRAC{3}{d^{\hbox to 2pt{\rm\tiny +\hss}}}{dt}}%
{\FRAC{3}{d^{\hbox to 2pt{\rm\tiny +\hss}}}{dt}}}}

\def\average#1,#2,{{1\over #2} \sum_{#1}^{#2}}

\def\eye(#1){{\bf(#1)}\quad}

\newtheorem{theorem}{{\bf Theorem}}

\def\eq#1/{(\ref{e:#1})}

\newcommand{\beqn}[1]{\notes{#1}%
\begin{eqnarray} \elabel{#1}}

\newcommand{\eeqn}{\end{eqnarray} }

\newcommand{\beq}[1]{\notes{#1}%
\begin{equation}\elabel{#1}}

\newcommand{\eeq}{\end{equation}}

\def\bdes{\begin{description}}
\def\edes{\end{description}}

\newcounter{rmnum}

\newcounter{anum}

{\end{list}}

\def\ass(#1:#2){(#1\ref{#1:#2})}

\def\ritem#1{
\item[{\sf \ass(\current_model:#1)}]
}

\newenvironment{recall-ass}[1]{%
\begin{description}
\def\current_model{#1}}{
\end{description}
}

%% file: macros.tex
\long\def\comment#1{}

\newfont{\bb}{msbm10 scaled 1100}

\newcommand{\av}{{\bf a}}

\newcommand{\cv}{{\bf c}}
\newcommand{\dv}{{\bf d}}

\newcommand{\fv}{{\bf f}}

\newcommand{\hv}{{\bf h}}

\newcommand{\qv}{{\bf q}}
\newcommand{\rv}{{\bf r}}

\newcommand{\uv}{{\bf u}}

\newcommand{\xv}{{\bf x}}
\newcommand{\yv}{{\bf y}}
\newcommand{\zv}{{\bf z}}

\newcommand{\Am}{{\bf A}}

\newcommand{\Dm}{{\bf D}}

\newcommand{\Fm}{{\bf F}}

\newcommand{\Xm}{{\bf X}}

\newcommand{\Ac}{{\cal A}}

\newcommand{\Fc}{{\cal F}}
\newcommand{\Gc}{{\cal G}}

\newcommand{\Nc}{{\cal N}}

\newcommand{\Xim}{\boldsymbol{\Xi}}

\renewcommand{\arg}{{\hbox{arg}}}

\newcommand{\SNR}{{\sf SNR}}

\newcommand{\transp}{{\sf T}}